# Vanishing nematic order beyond the pseudogap phase in overdoped cuprate superconductors

**Authors:** Naman K. Gupta[1], C. McMahon[1], R. Sutarto[2], T. Shi[1], R. Gong[1], Haofei I. Wei[3], K. M. Shen[3], F. He[2], Q. Ma[4], M. Dragomir[5,6], B. D. Gaulin[4,7], D. G. Hawthorn[1,*]

**Affiliations:**

[1]Department of Physics and Astronomy, University of Waterloo, Waterloo, Ontario N2L 3G1, Canada.

[2]Canadian Light Source, Saskatoon, Saskatchewan, S7N 2V3, Canada.

[3]Laboratory of Atomic and Solid State Physics, Department of Physics, Cornell University, Ithaca, NY 14853, USA.

[4]Department of Physics and Astronomy, McMaster University, Hamilton, ON L8S 4M1 Canada.

[5]Department of Chemistry and Chemical Biology, McMaster University, Hamilton, ON, L8S 4M1, Canada.

[6]Electronic Ceramics Department, Jožef Stefan Institute, 1000 Ljubljana, Slovenia.

[5]CIFAR, 661 University Ave., Toronto, ON, M5G 1M1, Canada.

[*]To whom correspondence should be addressed; **E-mail**: dhawthor@uwaterloo.ca

**Abstract:**

During the last decade, translational and rotational symmetry-breaking phases — density wave order and electronic nematicity — have been established as generic and distinct features of many correlated electron systems, including pnictide and cuprate superconductors. However, in cuprates, the relationship between these electronic symmetry-breaking phases and the enigmatic pseudogap phase remains unclear. Here we employ resonant x-ray scattering in a cuprate high-temperature superconductor $La_{1.6-x}Nd_{0.4}Sr_xCuO_4$ (Nd-LSCO) to navigate the cuprate phase diagram, probing the relationship between electronic nematicity of the Cu 3*d* orbitals, charge order, and the pseudogap phase as a function of doping. We find evidence for a considerable decrease in electronic nematicity beyond the pseudogap phase, either by raising the temperature through the pseudogap onset temperature *T*\* or increasing doping through the pseudogap critical point, *p*\*. These results establish a clear link between electronic nematicity, the pseudogap, and its associated quantum criticality in overdoped cuprates. Our findings anticipate that electronic nematicity may play a larger role in understanding the cuprate phase diagram than previously recognized, possibly having a crucial role in the phenomenology of the pseudogap phase.

**Significance statement:**

Understanding the character of the enigmatic pseudogap phase is a fundamental problem in the physics of the cuprate superconductors. Key to this problem is identifying what types of order, if any, are associated with pseudogap. Here we establish using resonant x-ray scattering that electronic nematicity – a rotational symmetry-breaking of the electronic structure – is linked to the onset of the pseudogap phase in the cuprate $La_{1.6-x}Nd_{0.4}Sr_xCuO_4$ (Nd-LSCO). We report vanishing of electronic nematicity either by raising the temperature through the onset of the pseudogap phase or by increasing hole-doping through the pseudogap critical point, indicating that electronic nematic order is intimately tied to the pseudogap.

**Main text:**

**Introduction**

There is a growing realization that the essential physics of the cuprate high-temperature superconductors, and perhaps other strongly correlated materials, involves a rich interplay between different electronic symmetry-breaking phases[1-3] like superconductivity, spin or charge density wave (SDW or CDW) order[4-7], anti-ferromagnetism, electronic nematicity[8-14], and possibly other orders such as pair density wave order[15] or orbital current order[16].

One or more of these orders may also be linked with the existence of a zero-temperature quantum critical point (QCP) in the superconducting state of the cuprates, similar to heavy-fermion, organic, pnictide, and iron-based superconductors[17-19]. The significance of the QCP point in describing the properties of the cuprates, as a generic organizing principle where quantum fluctuations in the vicinity of the QCP impact a wide swath of the cuprate phase diagram, remains an open question. Evidence for such a QCP and its influence include a linear in temperature resistivity extending to low temperature, strong mass enhancement via quantum oscillation studies[20], and an enhancement in the specific heat[21] in the field induced normal state, with some of the more direct evidence for a QCP in the cuprates coming from measurements in the material $La_{1.6-x}Nd_{0.4}Sr_xCuO_4$ (Nd-LSCO). Moreover, the QCP also appears to be the endpoint of the pseudogap phase[21] that is marked, among other features, by transition of the electronic structure from small Fermi surface that is folded or truncated by the antiferromagnetic zone boundary in the pseudogap phase to a large Fermi surface at higher doping[22,23] that is consistent with band structure calculations[24]. However, in the cuprates neither the QCP nor the change in the electronic structure have been definitively associated with a particular symmetry-breaking phase.

In this article, we interrogate the possibility that the cuprates exhibit a connection between electronic nematic order, the pseudogap and its associated QCP. In the pnictide superconductors, which are similar in many respects to the cuprates, electronic nematic order is more clearly established experimentally and there have been reports of nematic fluctuations[25], non-Fermi liquid transport[26] and a change in the topology of the Fermi surface associated with a nematic QCP[27]. Electronic nematicity refers to a breaking of rotational symmetry of the electronic structure in a manner that is not a straightforward result of crystalline symmetry, such that an additional electronic nematic order parameter, beyond the structure would be required to describe the resulting phase. The manifestation of nematic order may therefore depend on the details of the crystal structure of the materials, such as whether the structure is tetragonal or orthorhombic. However, such a state can be difficult to identify in materials that have orthorhombic structures, which would naturally couple to any electronic nematic order and vice versa. Despite these challenges, experimental evidence for electronic nematic order that is distinct from the crystal structure include reports of electronic nematicity from bulk transport[8-10] and magnetometry measurements[11] in YBCO, scanning tunneling microscopy (STM)[13,14,28] in $Bi_2Sr_2CaCu_2O_{8+\delta}$ (Bi2212), inelastic neutron scattering[12] in $YBa_2Cu_3O_y$ (YBCO), and resonant x-ray scattering[29] in $(La,Nd,Ba,Sr,Eu)_2CuO_4$. Moreover, STM studies in Bi2212 have reported intraunit cell nematicity disappearing around the pseudogap endpoint[30] which also seems to be a region of enhanced electronic nematic fluctuations[31,32]. In YBCO, there have also been reports of association between nematicity and the pseudogap onset temperature[9,11].

Here, we use resonant x-ray scattering to measure electronic nematic order in the cuprate $La_{1.6-x}Nd_{0.4}Sr_xCuO_4$ (Nd-LSCO) as a function of doping and temperature, to explore the relationship of electronic nematicity with the pseudogap phase. While evidence that a quantum critical point governs a wide swath of the phase diagram in hole-doped cuprates and is generic to many material systems remains unclear, investigation of Nd-LSCO provides the opportunity to probe the evolution of electronic nematicity over a wide range of doping in the same material system where some of the most compelling signatures of quantum criticality and electronic structure evolution have been observed. These include a divergence in the heat capacity[21], a change in the electronic structure from angle-dependent magnetoresistance (ADMR) measurements[24] in the vicinity of the QCP at $x = 0.23$, and the onset of the pseudogap[23]. Our main result is that we observe a vanishing of the electronic nematic order in Nd-LSCO as hole doping is either increased above $x = 0.23$, which has been identified as the QCP doping for this system[21], or when temperature is increased above the pseudogap onset temperature[23] $T^*$. These observations indicate that electronic nematicity in Nd-LSCO is intimately linked to the pseudogap phase.

**Resonant x-ray scattering on Nd-LSCO**

Our study uses resonant soft x-ray scattering (RSXS) measurements of the (001) Bragg peak to probe electronic nematicity in Nd-LSCO. Previous work by Achkar et al.[29] showed that the resonant x-ray scattering measurements of the (001) Bragg peak provide sensitivity to electronic nematicity. Understanding how this measurement reveals electronic nematicity requires examination of Nd-LSCO's crystal structure. Nd-LSCO exhibits a structural phase transition from the low temperature orthorhombic (LTO) to the low-temperature tetragonal phase (LTT) at temperatures between 60 and 90 K, depending on doping[33,34]. The LTO phase is characterized by tilting of the $CuO_6$ octahedra along an axis in the $CuO_2$ plane, but diagonal to the Cu-O bond. In the LTT phase, the axis by which the $CuO_6$ octahedra are tilted rotates 45° to be along either the *a* or *b* axes, such that each individual $CuO_2$ layer structurally breaks $C_4$ rotational symmetry as shown in Fig. 1. The axis of rotation of the octahedra, however, rotates by 90° between neighbouring layers producing a crystal structure that is tetragonal, despite having orthorhombic layers. Consequently, probes that average over the (neighbouring) layers would yield the same result along the *a* and *b* axes and may not reveal nematic order.

Resonant x-ray scattering of the (001) Bragg peak, however, provides a direct measure of the difference in the symmetry of the electronic structure between neighbouring layers[29,35]. Off resonance or with conventional x-ray diffraction, the (001) Bragg peak is forbidden, based on the ionic positions of atoms in the LTT or LTO phases. However, resonant x-ray scattering is sensitive to orbital symmetry, which differs for atoms in neighbouring layers in the LTT phase. Moreover, by tuning the photon energy to correspond to different atoms within the unit cell, the electronic symmetry-breaking of the $CuO_2$ planes — which contain the Cu *d* and O *p* states that cross the Fermi level $E_F$ and are most relevant to the low energy physics of the cuprates — can be differentiated from the $La_2O_2$ spacer layer, which do not have states that cross $E_F$ but would be sensitive to changes to their orbital symmetry induced purely by structural distortions.

Achkar et al.[29] showed that the temperature dependence of the (001) peak intensity exhibited a dichotomy when the photon energy was tuned to different atoms in the unit cell. An abrupt first-order-like increase at the LTO → LTT transition (consistent with previous measurements of the structural transition) was observed when the photon energy is at the La-$M$ edge (~826 eV) or an energy associated with apical oxygen (~532.3 eV) — atoms in the $La_2O_2$ spacer layer[29,35]. However, a more gradual temperature dependence was observed when the photon energy is tuned to the Cu-$L$ edge (~931.5 eV) or an energy associated with in-plane O (~528.5 eV) — atoms in the $CuO_2$ planes[29]. The different temperature dependencies can be understood in terms of a metanematic order, in analogy to a metamagnetic order in magnetism. Here electronic nematic order in the $CuO_2$ planes has a temperature dependent susceptibility and couples to the lattice order. This provides an additional nematic contribution to the (001) structural Bragg peak. Effectively this is akin to an internal strain applied to individual $CuO_2$ planes by the lattice distortion, with the electronic structure of the $CuO_2$ planes (probed by the (001) peak intensity at the Cu-$L$ edge) exhibiting a response that depends on the electronic nematic susceptibility. Importantly, Achkar et al. also showed that this electronic nematic order is coupled to but distinct from charge density wave order. Here we investigate electronic nematicity as a function of doping (through the pseudogap QCP, $p^*$) and temperature (through the pseudogap onset, $T^*$). Since our measure of electronic nematicity requires the LTT phase, the latter measurement requires a sample with $T_{\text{LTT}} > T^*$, which is achieved here at a doping level of $x = 0.225$.

**Results:**

**Doping evolution of electronic nematicity**

In Fig. 2, we show the intensity of the (001) peak as a function of temperature ($T$) at the Cu-$L$ edge (probing the $CuO_2$ planes) and the O-$K$ edge at an energy sensitive to apical O (probing the $La_2O_2$ layer). Note, here the data at the different edges is arbitrarily scaled to match in intensity below the LTO → LTT phase transition to highlight the difference in temperature dependence for different doping levels. The onset temperature of electronic nematicity, $T_{EN}$, is identified as the temperature where the (001) peak first exhibits a notable deviation between Cu-$L$ and apical O-$K$ edges. In Fig. 2A data at $x = 0.125$ from Ref. (29) is reproduced, which showed a distinct difference in the $T$ dependence of the (001) peak for measurements at the Cu-$L$ and O-$K$ edges. As the doping is increased in Nd-LSCO, a similar difference in the peak intensity between Cu-$L$ and O-$K$ edges is observed for $x = 0.17, 0.18, 0.19$ and $0.225$ (see Fig. 2A to 2E). However, for $x = 0.24$, as the hole doping is increased across the pseudogap QCP (at $x \sim 0.23$), within the measurement accuracy we detect no difference in the $T$ dependence of the (001) peak when measured at the apical O-$K$ and Cu-$L$ edges (see Fig. 2F).

Notably we also observe that for $x = 0.225$, within the measurement accuracy, we detect no difference in the $T$ dependence of the (001) peak at the apical O-$K$ and Cu-$L$ edges above the pseudogap onset temperature $T^*$, but a distinct difference in temperature dependence below $T^*$ ($T^*$ is identified in Ref. (23) by the resistivity upturn at 50 K in $x = 0.22$ and 40 K in $x = 0.23$). Our principal conclusion from this result is that static, long-range ordered electronic nematicity in Nd-LSCO is a characteristic of the pseudogap phase, vanishing either upon traversing the QCP with increasing doping or by increasing the temperature above $T^*$.

**Doping evolution of CDW order**

We also investigated the presence of CDW order in the same series of samples to identify the temperature and doping extent of the CDW phase in the cuprate phase diagram. We observe CDW order in $x = 0.125$ and $x = 0.17$ samples, as shown in Fig. 3, but not for $x = 0.18, 0.19$, or 0.24 above the base temperature (~22 K) of the measurements (see Supplementary Information Fig. S4). The onset temperature of CDW order is consistent with the recent Seebeck coefficient measurement[36] in Nd-LSCO. In contrast, SDW order has been observed at low temperature up to a doping as high as $x = 0.26$ by elastic neutron scattering. However, the onset of SDW order ($T_{SDW} = 35$ K in $x = 0.19$) occurs at lower temperatures than our observations of electronic nematicity[37]. This shows that the existence of stripe-like CDW or SDW order, that breaks both translational and rotational symmetry, does not appear to be the solely responsible for the electronic nematicity. Rather, electronic nematicity is either a distinct order or a melted stripe phase that retains rotational symmetry-breaking but not the translational symmetry-breaking of the CDW or SDW order[38]. A schematic phase diagram for Nd-LSCO that summarizes the doping evolution of the LTO $\rightarrow$ LTT phase transition, the onset of CDW order, and our evidence for long-range ordered electronic nematicity, with onset temperature $T_{EN}$, is depicted in Fig. 4.

Notably, there has been a recent report in the related material LSCO[39], which remains in the LTO phase, that CDW order persists to higher doping levels than in Nd-LSCO; with CDW observed up to at least $p = 0.21$ – which is past the point where the Fermi surface transitions from hole-like to electron-like.

**Discussion**

Our observation that electronic nematic order in Nd-LSCO is associated with the pseudogap phase is reinforced by recent measurements in other cuprates families that do not exhibit the LTT structural phase, indicating that an electronic nematic phase terminating at the QCP in overdoped cuprates may be a generic property of the cuprate phase diagram. Specifically in Bi2212, low-temperature STM measurements indicate a disappearance of nematicity at the pseudogap QCP[30] and recent elasto-resistance[31] and Raman scattering[32] measurements have shown dynamical electronic nematic fluctuations that are enhanced — in the overdoped region between $p = 0.20$ and 0.23 — and nematic quantum criticality which is consistent with the putative quantum critical point for Bi2212 at $p^* \sim 0.22$.

The questions remain whether electronic nematicity is fundamentally linked to the origins of the pseudogap or is simply stabilized when the pseudogap onsets, and what role if any quantum nematic fluctuations play in the physics of the cuprates. In the proximity of $p^*$, it has also been established that the electronic structure of the cuprates undergoes a Lifshitz transition from a small Fermi surface or Fermi arcs in the pseudogap phase to a large hole like Fermi surface at higher doping in Nd-LSCO[23,24,40] and other cuprate families[30,22,41] Perhaps most notably, in Bi2212[14] the Lifshitz transition was shown to occur simultaneous with the disappearance of nematicity, similar to what we observe in Nd-LSCO. This change in the nature of the Fermi surface may provide the conditions for the presence or absence of electronic nematicity. Or, as suggested by recent theoretical proposals[42], the doping dependence of the electronic nematicity observed here may account for the change in the Hall number around the Lifshitz transition.

It is also possible that the nematicity observed here represents a "vestigial nematicity" that is connected to the pseudogap phase. It has been recently highlighted that in contrast to a CDW order, which can open a gap in the Fermi surface, electronic nematic order is a $q_{x,y} = 0$ order and is not expected to open a gap in the Fermi surface[43,44]. As such, one would not anticipate a natural connection between the pseudogap, which is evidenced by the opening of an antinodal gap, and electronic nematicity. However, a "vestigial nematic" state – derived from disordering a unidirectional CDW, but lacking long range order, may open up a gap and is plausibly connected to the pseudogap phase[44]. Such vestigial nematic state is consistent with our measurements, where nematicity is present in the pseudogap phase, but also at dopings and temperatures where CDW order is not observed.

**Conclusion**

In this study, we report the evidence for vanishing electronic nematic order as the doping is increased beyond the pseudogap QCP at $x = 0.23$, that has been identified in Nd-LSCO, or when the temperature is increased above the pseudogap onset, $T^*$. This suggests that nematicity is associated with the pseudogap and may possibly be the fluctuating order responsible for the quantum critical phenomena near the pseudogap QCP. Consequently, nematicity may play a larger role in understanding the cuprate phase diagram that has been previously recognized, possibly having a key role in the nature of the pseudogap phase.

**Materials and Methods**

Resonant soft x-ray scattering measurements and X-ray absorption spectroscopy measurements at the Cu-*L* and O-*K* edges were performed using the in-vacuum four-circle diffractometer at the Canadian Light Source's REIXS beam line[45]. Measurements were performed on single-crystal samples of $La_{1.6-x}Nd_{0.4}Sr_xCuO_4$ (Nd-LSCO) with $x$ = 0.17, 0.18, 0.19, 0.225 and 0.24. Details of sample growth and characterization of the other samples can be found in Ref. (34). Data on the $x$ = 0.125 sample is reproduced from Ref. (29). All measurements were performed on the ⟨001⟩ sample surface, with the samples cleaved in air prior to insertion into the vacuum chamber. The vacuum chamber pressure for all measurements was ~$2\times10^{-10}$ Torr. The background-subtracted data was fitted to Lorentzian line shapes – raw data, analysis, details, and supporting text are described for the $x$ = 0.17 and $x$ = 0.24 Nd-LSCO single crystal sample in the Supplementary Information. The (001) Bragg peak scattering intensity with temperature is obtained in a similar fashion for other doping levels. The crystallographic orientation of the samples was verified in the diffractometer using the (0 0 2) and (±1 0 3) structural Bragg peaks measured with photon energy >2 keV. Reciprocal lattice units (r. l. u.) were defined using the lattice constants as $a = b$ = 3.787 Å and $c$ =13.24 Å for Nd-LSCO, where the *a* and *b* axes are referenced to the high temperature tetragonal (HTT) unit cell[34]. For x > 0.15, the LTO to LTT transition is between LTO (*Bmab* space group) and LTT ($P4_2/ncm$ space group). For the $x$ = 0.125 sample, the LTO to LTT phase transition is mediated by the LTO2 phase (*Pccn* space group), with reduced orthorhombicity relative to LTO and a tilt pattern that alternates between neighbouring layers, like the LTT phase[34]. The transition depicted in Fig. 2A at 70 K is between the LTO and LTO2 phases, but is referred to LTT in Fig. 2A and at $x$ = 0.125 in Fig. 4 for simplicity.

**Acknowledgements:** The authors acknowledge insightful discussions with S. A. Kivelson, M. J. P. Gingras and L. Taillefer. **Funding:** This work was supported by the Natural Sciences and Engineering Research Council of Canada (NSERC), Air Force Office of Scientific Research grants FA9550-15-1-0474 and FA9550-21-1-0168, and the Canada First Research Excellence Fund, Transformative Quantum Technologies Program. Research described in this paper was performed at the Canadian Light Source, a national research facility of the University of Saskatchewan, which is funded by the Canada Foundation for Innovation (CFI), the NSERC, the National Research Council Canada (NRC), the Canadian Institutes of Health Research (CIHR), the Government of Saskatchewan, and the University of Saskatchewan. B.D.G. acknowledges support from the CIFAR as a CIFAR Fellow.

**Author contributions:** Conceptualization: DGH**,** Formal Analysis: NKG, CM, TS and DGH**,** Funding acquisition: KMS, BDG and DGH**,** Investigation: NKG, CM, RS, TS, RG, HIW and DGH**,** Project administration: DGH**,** Resources: RS and FH, Materials synthesis and crystal growth: QM, MD and BDG**,** Supervision: KMS, BDG, and DGH**,** Writing – original draft: NKG and DGH**,** Writing – review & editing: RS, NKG, DGH, MD, KMS and BDG.

**Competing interests:** Authors declare no competing interests.

**Data and materials availability:** All study data are included in the article and/or Supplementary information.

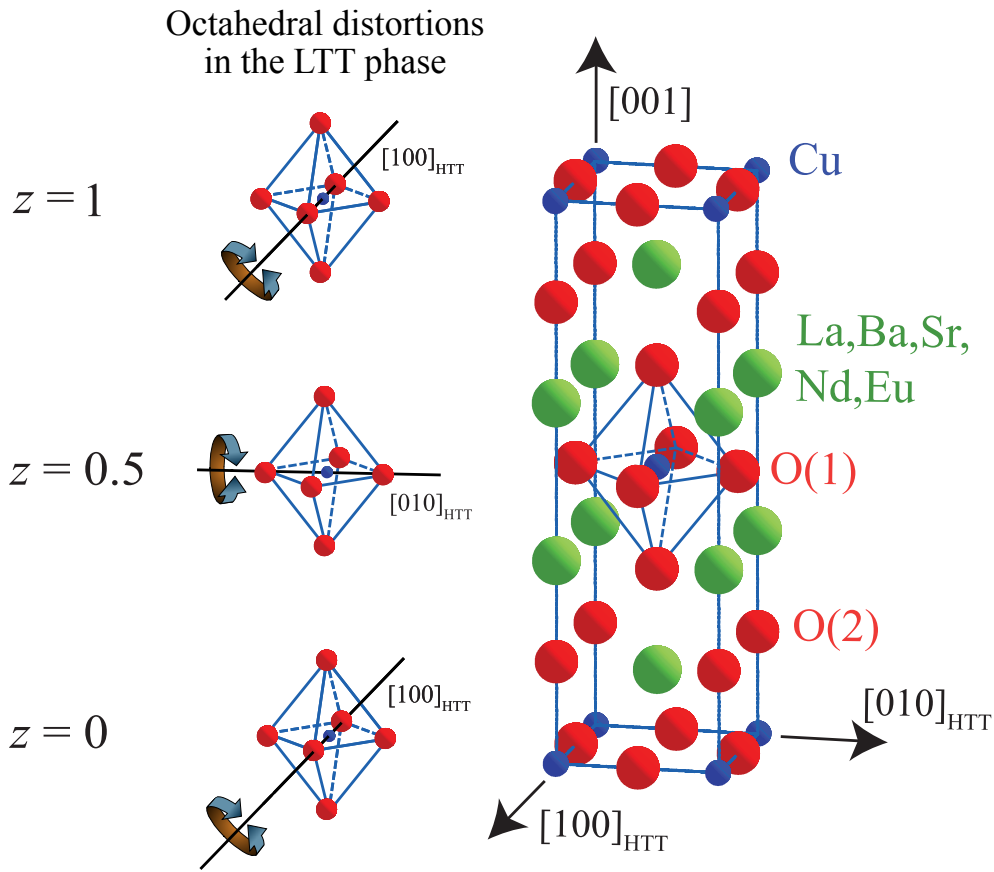

**Figure 1. Crystal structure of Nd-LSCO. Right:** Unit cell of $La_{1.6-x}Nd_{0.4}Sr_xCuO_4$ in the high-temperature tetragonal phase (HTT). O(1) and O(2) are in-plane and apical oxygen sites, respectively. In the HTT phase, the axes of the $CuO_6$ octahedra are aligned with the axes of the unit cell. The low-temperature orthorhombic (LTO) and low-temperature tetragonal (LTT) phases are understood in terms of rotations of the $CuO_6$ octahedra. **Left:** In the LTT phase, octahedra rotate about axes parallel to the Cu-O bonds, with the rotation axis of the octahedra alternating between the *a* and *b* axes of neighboring planes (z = 0 and z = 0.5). This induces $C_4$ symmetry-breaking of the average (or intra-unit cell) electronic structure within an individual $CuO_2$ plane, referred to here as electronic nematicity.

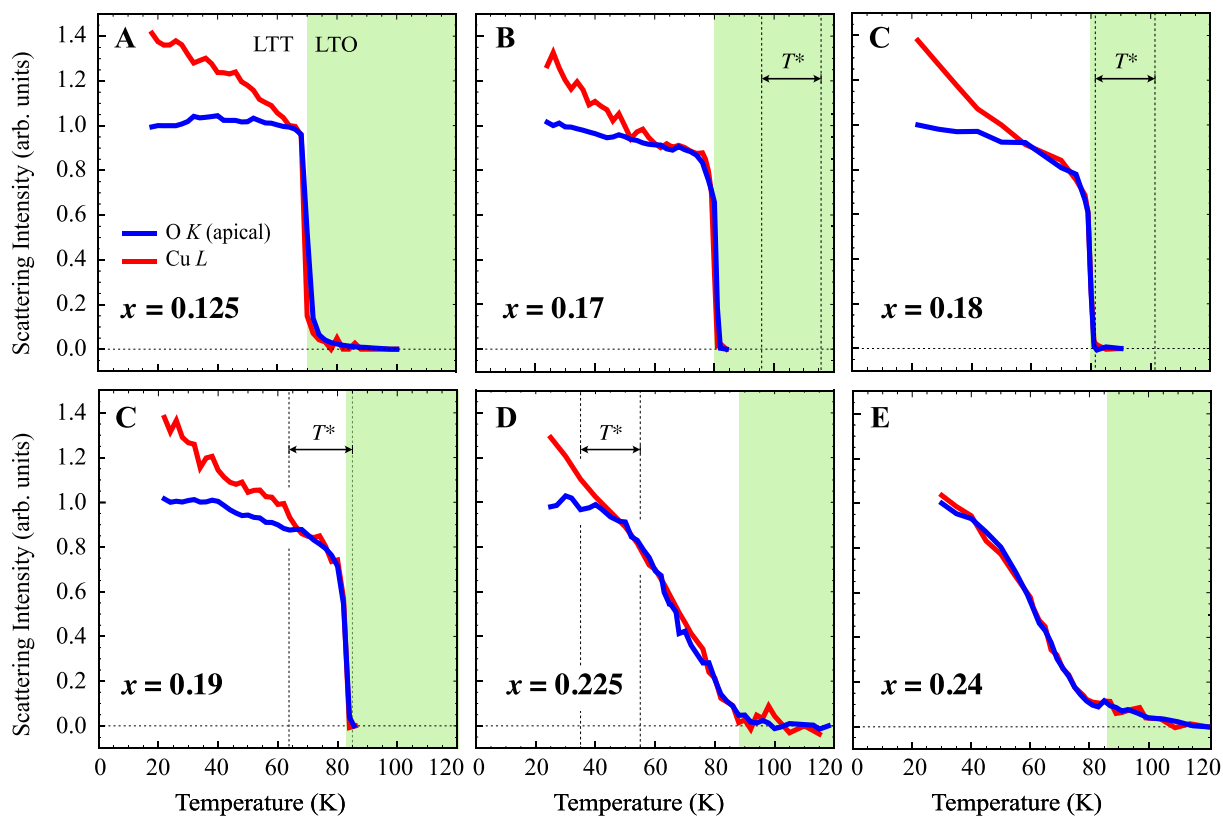

**Figure 2. Doping and temperature dependence of the (001) Bragg peak intensity in Nd-LSCO.** The intensities of the (001) Bragg peak as a function of temperature with photon energy tuned to the apical O-*K* (blue) and Cu-*L* (red) edges for $La_{1.6-x}Nd_{0.4}Sr_xCuO_4$ with Sr doping levels: *x* = 0.125, 0.17, 0.18, 0.19, 0.225 and 0.24. The white and green region indicate the LTT phase and the LTO phase respectively. The O-*K* and Cu-*L* data are arbitrarily scaled to match in intensity below the LTT phase transition in order to highlight differences in temperature dependence. The dotted lines mark the pseudogap onset temperature, *T\**, in Nd-LSCO based on the high field electrical resistivity measurements from Ref. (23). Data on the *x* = 0.125 sample is reproduced from Ref. (29).

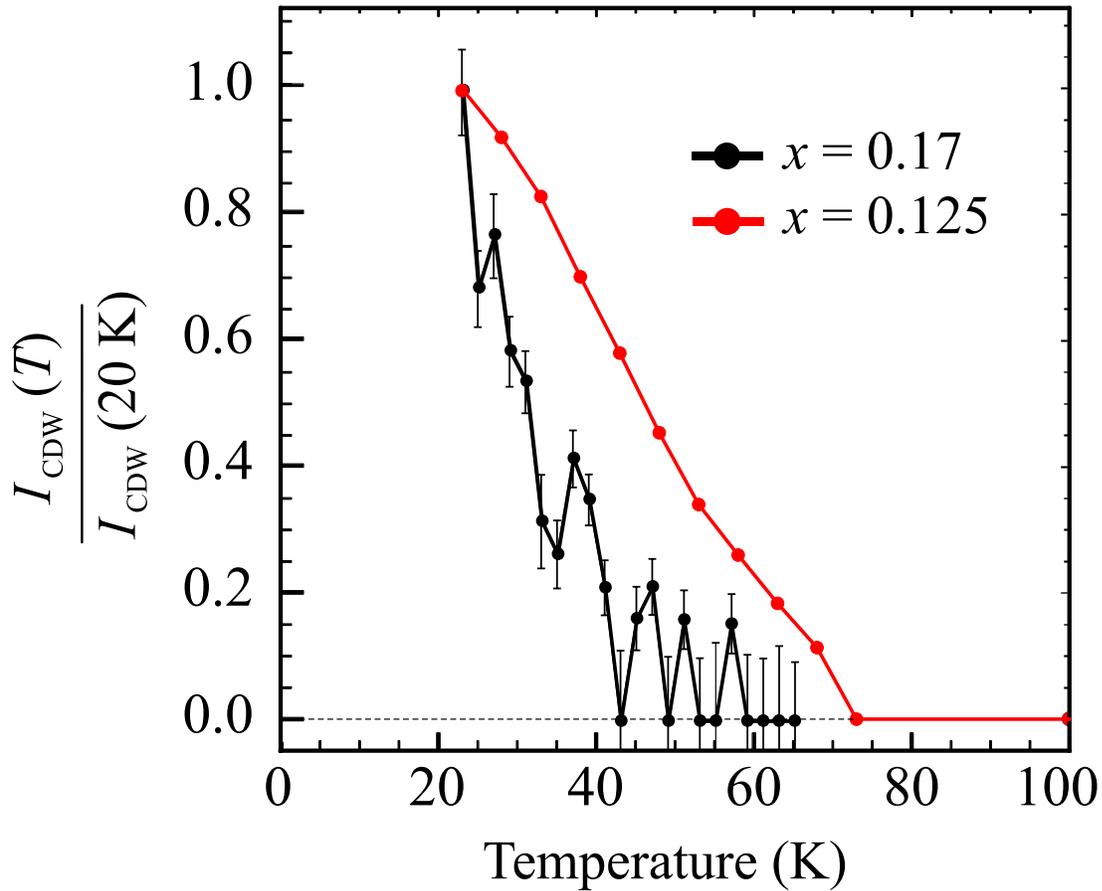

**Figure 3. Doping and temperature dependence of the CDW peak intensity.**

The intensities for the CDW peak at Cu-*L* edge (normalized low temperature value ∼ 24 K) for $\text{La}_{1.6-x}\text{Nd}_{0.4}\text{Sr}_x\text{CuO}_4$ (Nd-LSCO) with Sr doping levels: *x* = 0.125 and 0.17. We did not observe a CDW signal at *x* = 0.18, 0.19 or 0.24 above our base temperature. Data on the *x* = 0.125 sample is reproduced from Ref. (29).

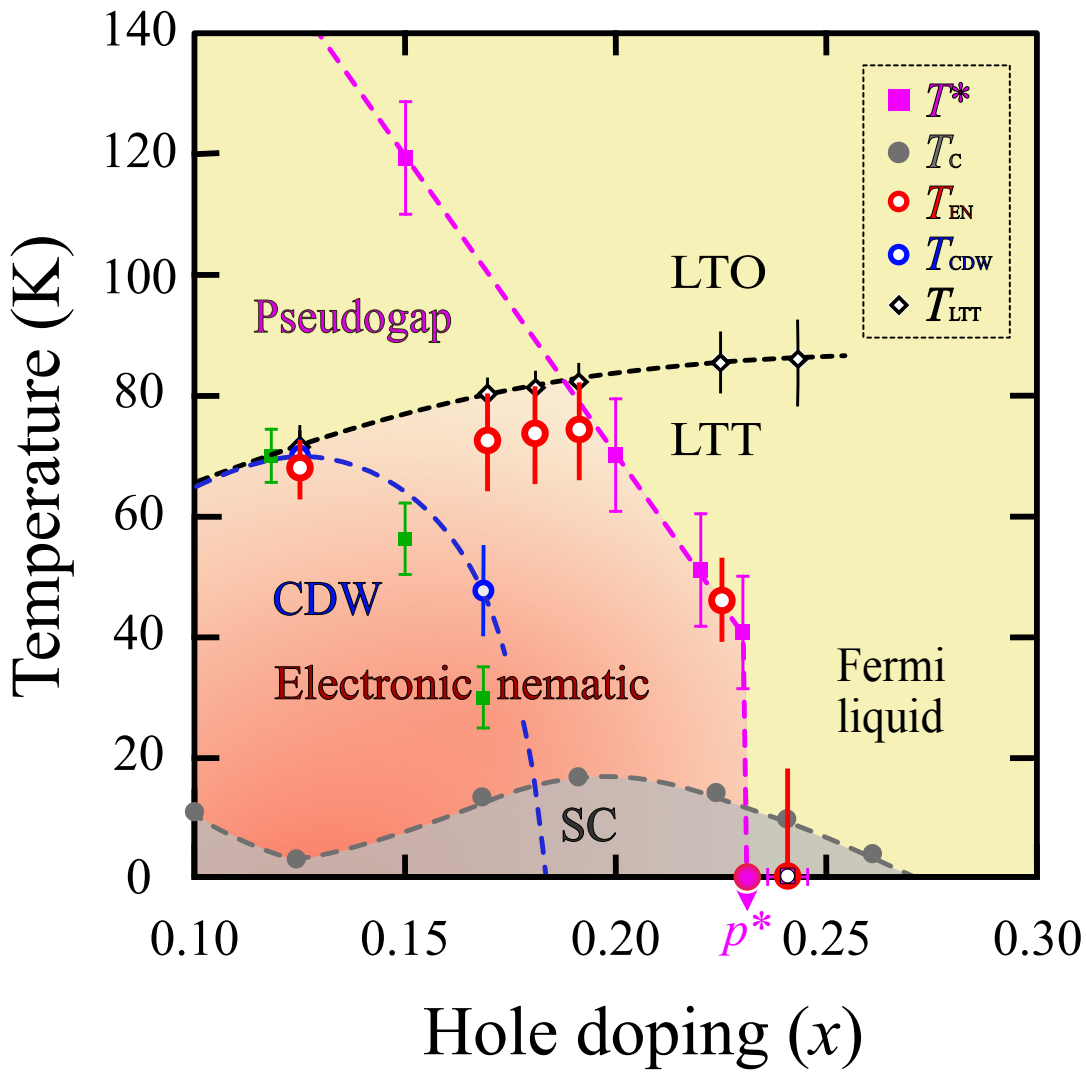

**Figure 4. Phase diagram of Nd-LSCO.**

The doping-temperature phase diagram of Nd-LSCO. The structural phase transition from the LTO to LTT phase, $T_{LTT}$ (black diamonds). The onset temperature of the charge density wave (CDW) order, $T_{CDW}$, using our resonant X-ray scattering measurement (blue circles) and recent Seebeck coefficient[36] measurement (green squares). The red circles mark the onset of the electronic nematic order in the LTT phase, $T_{EN}$. The superconducting (SC) transition temperature, $T_c$ (grey circles), has been reported in Ref. (34) on the same series of samples as used for our RSXS measurements. The existence of a quantum critical point at $p^* \sim 0.23$ has been identified in Ref. (21). The pseudogap onset temperature, $T^*$ (purple squares), as a function of hole doping is measured via the onset of the upturn in electrical resistivity $\rho(T)$ in Ref. (23). The black, purple, blue and grey dotted lines are a guide to the eye.

# Vanishing nematic order beyond the pseudogap phase in overdoped cuprate superconductors

**Authors:** Naman K. Gupta[1], Christopher McMahon[1], Ronny Sutarto[2], Tianyu Shi[1], Rantong Gong[1], Haofei I. Wei[3], Kyle M. Shen[3], Feizhou He[2], Qianli Ma[4], Mirela Dragomir[5,6], Bruce D. Gaulin[4,7], David G. Hawthorn [1,*]

*To whom correspondence should be addressed; **E-mail**: dhawthor@uwaterloo.ca

**This PDF file includes:**

Supplementary text

Figures S1 to S5

SI References

**Measurement and Temperature dependence of the (001) Bragg peaks**

Here we present representative measurements of the (001) peak and its temperature dependence in Nd-LSCO at doping level $x = 0.17$. Raw data, as shown in Fig. S1, in this sample is collected by a photodiode (PD) detector. This detector integrates over the incident photon energies, which includes both 1st order and a subdominant 2nd order (at twice the desired photon energy) contribution from the beamline monochromator. This 2nd order light contribution enables the measurement of the strong (002) structural Bragg peak of Nd-LSCO at angles that coincide with the detection of the (001) peak with 1st order light. Note, in Fig. S1, the $L$ value in reciprocal lattice vectors from Bragg's law is shown. However, due to refraction, the detected peak angles are different from the expected peak[28] at (001). Moreover, due to refraction, the (001) and (002) peaks are peaked at different angles or $L$ values.

Upon heating from the LTT to LTO phase, the (001) Bragg peak vanishes – which is around 81 K for the $x = 0.17$ sample – and only the $T$-independent structural (002) Bragg peak remains. Contributions from the (002) structural Bragg peak, can be subtracted from the total intensity (001) to yield the (001) peak intensities. This is achieved by subtracting the high temperature scans (in the LTO phase) from the total intensity.

The resulting (001) peaks at selected temperatures for both the apical O-$K$ edge (~533 eV) and Cu-$L$ edge (~931.5 eV) is shown in Fig. S2, along with Lorentzian fits.

Lorentzian lineshape profile fit the data well and exhibit a full width at half maximum (FWHM) that is temperature independent, as shown in Fig. S2(c). This is consistent with the (001) peak width analysis in Supplementary Information of Ref. (28) for Nd-LSCO at $x = 0.125$, which discusses that the FWHM is determined by x-ray absorption length and not the correlation length of electronic nematicity. Similar temperature independent peak widths are seen at other absorption edges and doping levels.

The $T$-independent FWHM implies that the $T$ dependence of the peak intensity and the integrated intensity is equivalent for the (001) peak. Scattering intensity plotted in Fig. S2(d) is obtained by the fitted-peak intensity (amplitude of the Lorentzian line shape) for the apical O-$K$ (blue curve) and Cu-$L$ (red curve) edge. Note that the data at the Cu-$L$ and apical O-$K$ edges in Fig. S2(d) is arbitrarily scaled to match in intensity below the LTO $\rightarrow$ LTT phase transition to highlight the difference in temperature dependence at the two transition edges.

For $x = 0.24$, we use an energy resolved Silicon Drift Detector (SDD) which can differentiate between the (001) and (002) peak contributions unlike the PD. Dataset and Lorentzian fits for this sample are shown in Fig. S3. Note that due to saturation of the energy resolved detector by the 2$^{nd}$ order light ($E_i \sim 1066$ eV) at apical O-$K$, contribution from associated $L$-values have been clipped out in determining the fitting parameters as shown in Fig. S3(b).

**Temperature dependence of the CDW peaks in Nd-LSCO**

The CDW order is evidenced by a broad and weak peak above the fluorescent background signal in resonant scattering around the (CDW) peak maximum[1,28] ~ ($\delta$ 0 1.5). The temperature dependence of the CDW peaks at the Cu-$L$ edge were measured by performing scans of the sample angle, at a fixed value of the detector angle for samples at Sr doping level of $x$ = 0.17, 0.18, 0.19 and 0.24. These scans vary both $H$ and $L$. However, since the CDW peaks are broad in $L$ these scans primarily identify the peak and its width in $H$. We detect CDW order for $x$ = 0.17 as shown in Fig. S4(a). A cubic background is subtracted from the data, and the resulting CDW peak are fit to Lorentzian line shapes, as shown in Fig. S4(b). Scattering intensity was obtained by measuring the fitted-peak amplitude at the Cu-$L$ edge (shown in Fig. 3 of the main text). However, we did not observe signatures of CDW order at $x$ = 0.18, 0.19 and 0.24 above our base temperature as shown in Fig. S4(c) and 4(d).

**X-ray absorption (XAS) for Nd-LSCO at O-$K$ and Cu-$L$ edge**

X-ray absorption (XAS) is measured as a function of photon energies – at the Cu-$L$ and O-$K$ edges – in Nd-LSCO at $x$ = 0.17 using total fluorescence yield (TFY), as shown in Fig. S5. At ~931.5 eV, where Cu-$L$ RSXS measurements in this study are performed, the XAS signal is dominated by the in-plane Cu 3$d$ states[44]. Moreover, at ~528.5 eV, the XAS signal is sensitive to the O 2$p$ states of the in-plane, O(1), states[45]. However, at

~532.3 eV, where O-$K$ edge RSXS measurements in this study are performed, the XAS signal shows sensitivity to the apical oxygen, O(2), the states in the spacer layer which are hybridized with (La, M) states[34,46]. The XAS measurements are qualitatively consistent between all the measured samples from $x = 0.125$ to 0.24.

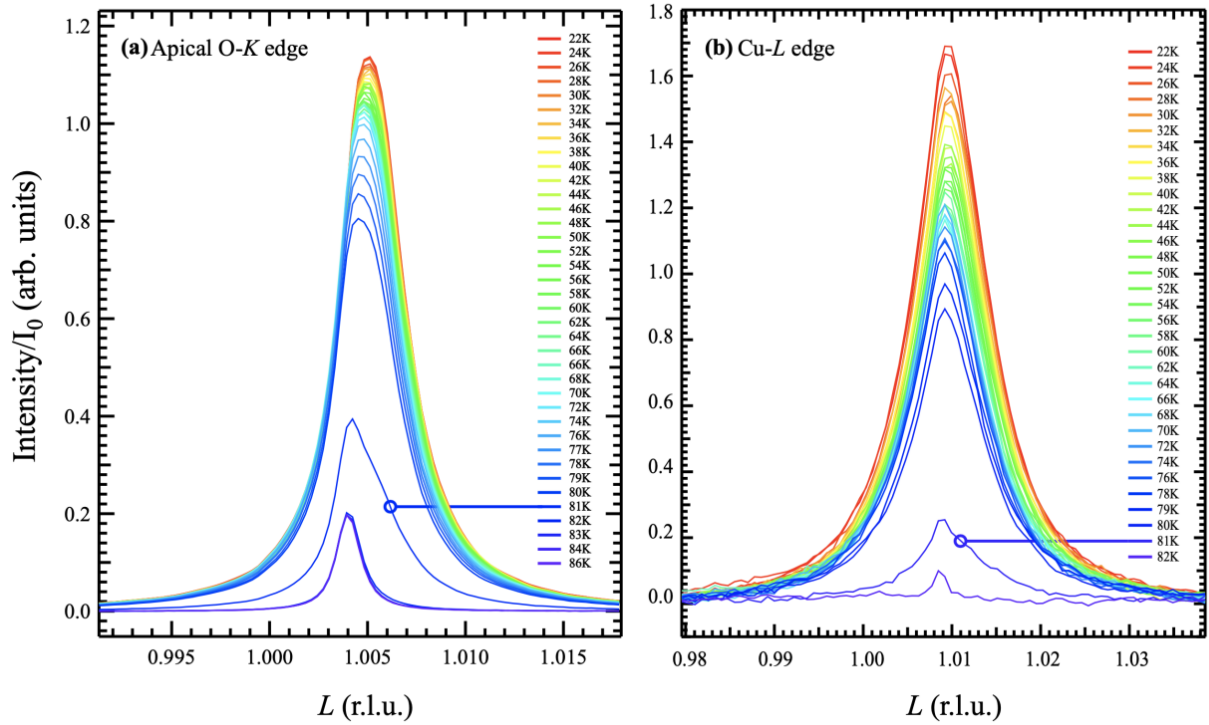

**Figure S1. Raw dataset for (001) Bragg peaks as a function of temperature for Nd-LSCO at *x* = 0.17 using a photodiode detector.** Recorded (001) peak intensity and shape as function of *L* at various temperatures at (a) apical O-*K* edge (~ 532 eV) and (b) in-plane Cu-*L* edge (~931.5 eV). The sharp (001) signal-intensity drop at ~81 K marks the LTT → LTO phase transition, above which the (001) peak vanishes and the photodiode only detects a weak, *T*-independent (002) structural Bragg peak coming from 2$^{nd}$ order light.

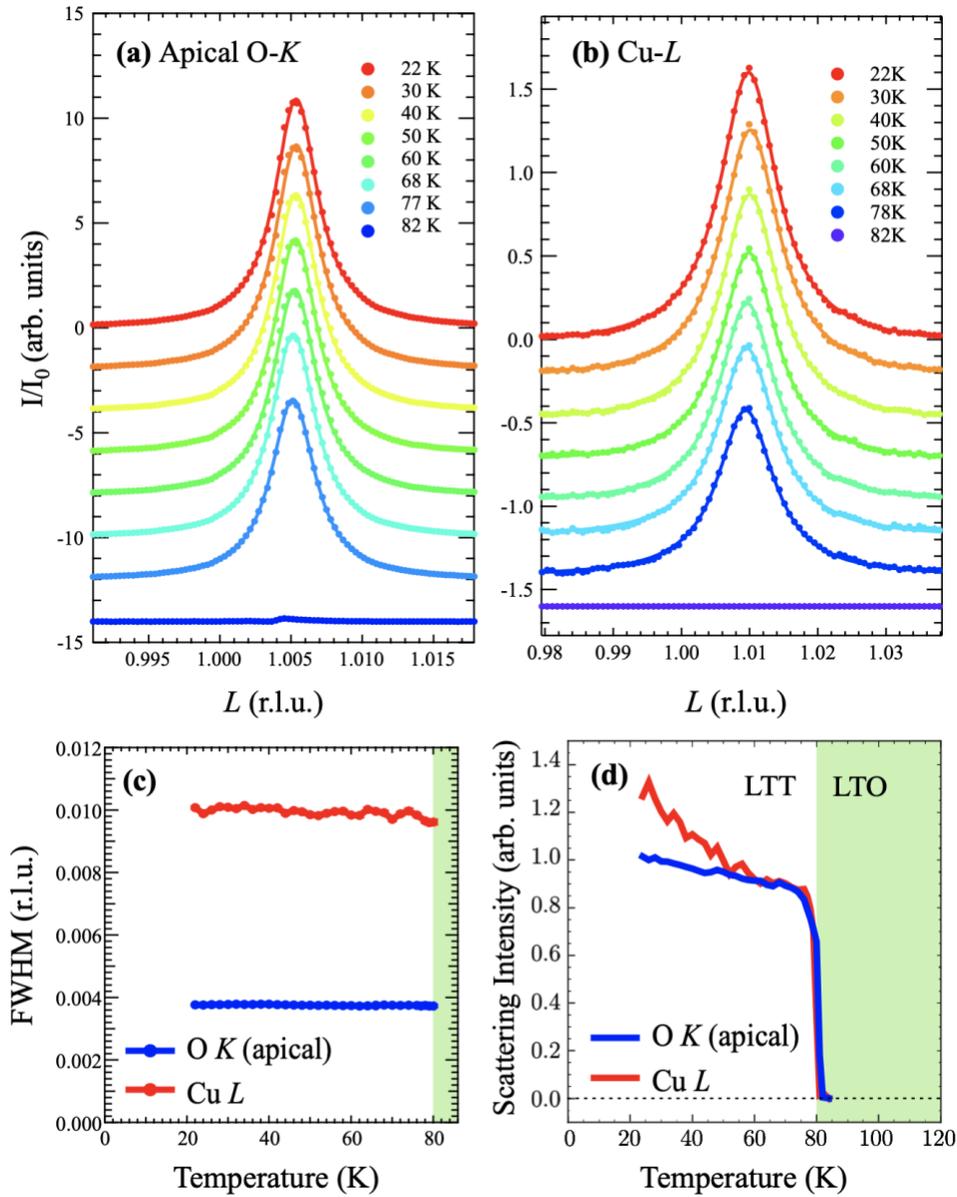

**Figure S2. Temperature dependence of (001) Bragg peaks in Nd-LSCO at $x$ = 0.17.**
Temperature dependence of the (001) Bragg peaks at **(a)** apical O-$K$ edge (~532 eV) and **(b)** in-plane Cu-$L$ edge (~931.5 eV). Solid lines are Lorentzian fits to the data (filled circles) at different temperatures. The peak intensity decreases with increasing temperature, but the width remains constant. **(c)** The *T* dependence of the FWHM of the Lorentzian fits for the (001) peak at the apical O-*K* and Cu-*L* edges. **(d)** The scattering intensity (peak amplitude of the Lorentzian fits) for both the edges. The intensities for apical O-*K* signal are normalized by the corresponding low-temperature values, $I_{O(001)} \sim 24\ \mathrm{K}$, with the intensity for Cu-*L* edge scaled to match the O-*K* intensity at a temperature below the LTO → LTT transition. The white and green region indicate the LTT phase and the LTO phase respectively in (c) and (d).

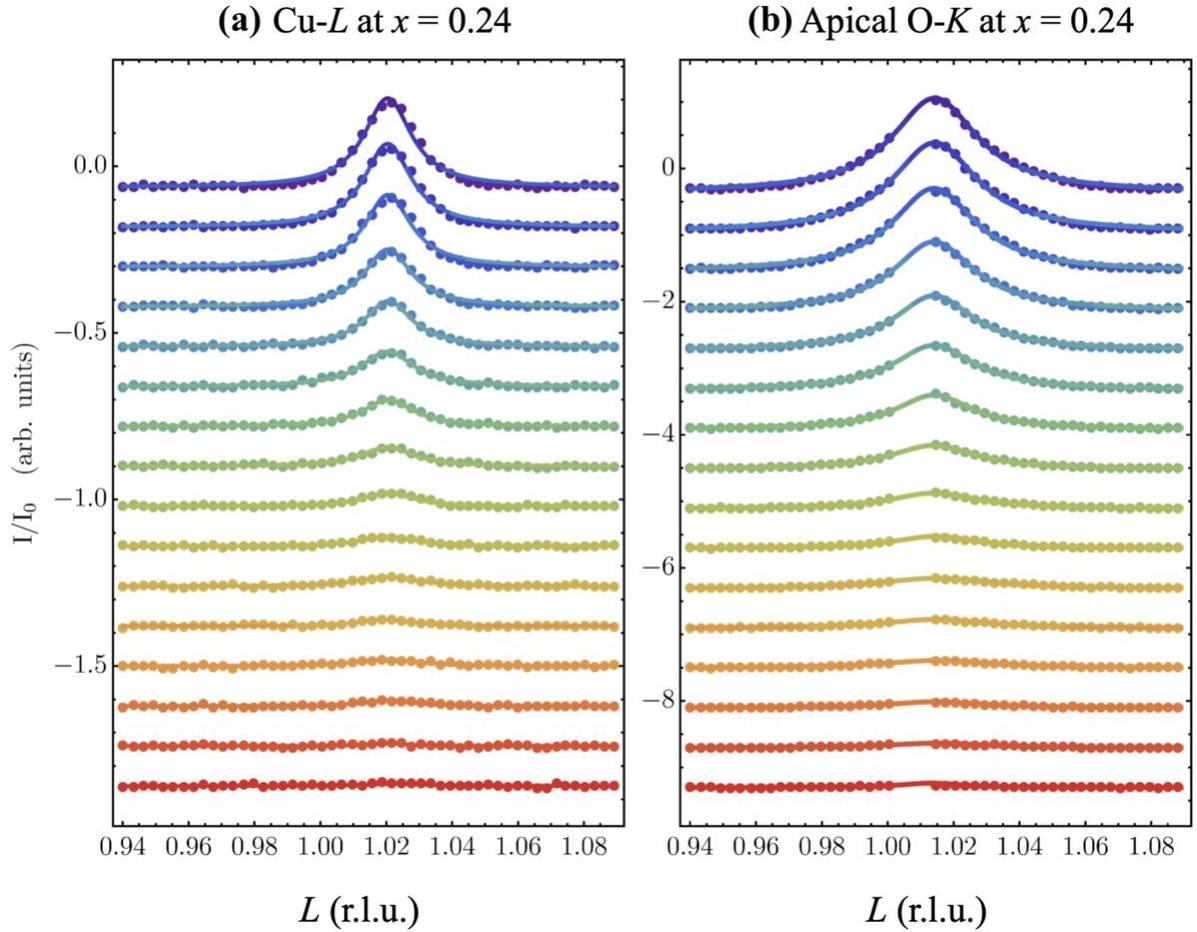

**Figure S3. Temperature dependence of (001) Bragg peaks in Nd-LSCO at *x* = 0.24.**
Temperature dependence of the (001) Bragg peaks for *x* = 0.24 sample at **(a)** apical O-*K* edge (~532 eV) and **(b)** in-plane Cu-*L* edge (~931.5 eV). The peak intensity decreases with increasing temperature, but the width remains constant. Solid lines are Lorentzian fits to the data (filled circles) at different temperatures. The (001) peak intensity is recorded by tracking the amplitude of the Lorentzian fits, see Fig. 2F in the main text.

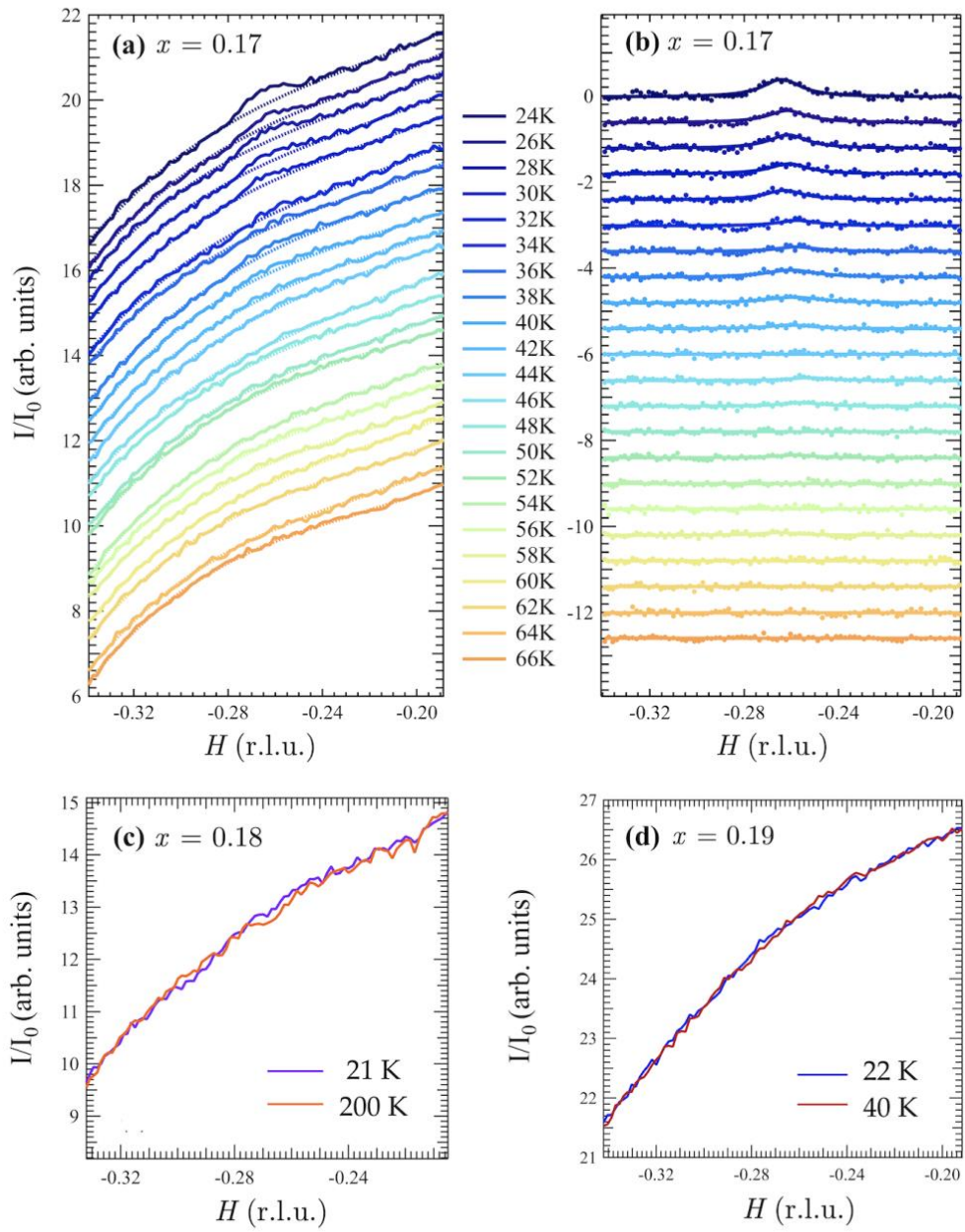

**Figure S4. Temperature dependence of the CDW peaks in Nd-LSCO. (a)** Measured intensity for scans as function of *H* at various temperatures through the **Q** = (-0.26, 0, 1.5) CDW peak at *x* = 0.17 at various temperatures. A cubic polynomial (dashed curve) is fit to the fluorescent background and subsequently subtracted from the data. **(b)** The polynomial background-subtracted data for $x$ = 0.17. Solid lines are Lorentzian fits to the background-subtracted data (filled circles). In both (a) and (b) the data at different temperatures are offset for clarity. **(c)** and **(d)** Measured intensity through **Q** = (-0.26, 0, 1.5) for (c) $x$ = 0.18 and (d) $x$ = 0.19 at high temperature and at our base temperature (~22K) showing the absence of measurable CDW order in these samples.

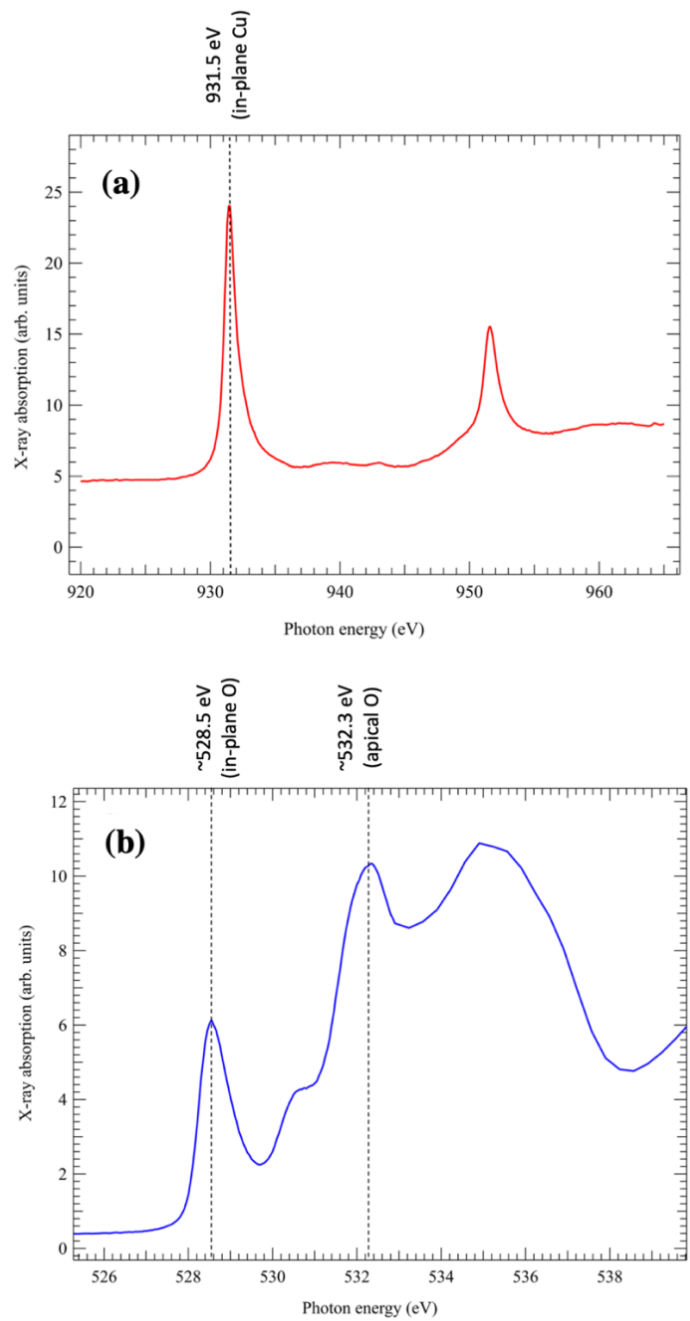

**Figure S5. X-ray absorption (XAS) at the Cu-*L* and O-*K* edges in Nd-LSCO.** The x-ray absorption (XAS) as a function of photon energy at the **(a)** Cu-*L* and **(b)** O-*K* edges in Nd-LSCO at $x$ = 0.17 measured using total fluorescence yield (TFY).